\documentclass[letterpaper, 10 pt, conference]{ieeeconf} 

\usepackage{amsfonts,amssymb,amsmath}
\usepackage{graphicx,graphics,epsfig,color}
\usepackage{rgsMacros}
\usepackage{subcaption}
\usepackage{comment}
\usepackage{hyperref}

\let\labelindent\relax
\usepackage{enumitem}
\usepackage{balance}

\definecolor{olivegreen}{rgb}{0.14,0.29,0}

\IEEEoverridecommandlockouts                              
\newtheorem{exe}{Example}
\newtheorem{corol}{Corollary}
\newtheorem{ass}{Assumption}
\newtheorem{proper}{Property}
\newtheorem{defin}{Definition}
\newtheorem{prob}{Problem}
\newtheorem{cla}{Claim}
\newtheorem{rem}{Remark}
\newtheorem{lem}{Lemma}
\newtheorem{prop}{Proposition}
\newtheorem{thm}{Theorem}
\newtheorem{fct}{Fact}

\newenvironment{example}{\begin{exe}}{\hfill $\square$ \end{exe}}
\newenvironment{remark}{\begin{rem} \rm}{ \end{rem}}

\newenvironment{theorem}{\begin{thm}}{\hfill $\square$ \end{thm}}

\newif\ifitsdraft

\newtheorem{dwellt}{Condition}

\newif\ifitsdraft

\usepackage{mathtools}
\definecolor{cadmiumgreen}{rgb}{0.0, 0.42, 0.24}
\usepackage{tikz}
\usetikzlibrary{shapes,positioning,intersections,quotes}
\usepackage{pgfplots}
\pgfplotsset{compat=1.16}

\tikzset{
    cross/.pic = {
    \draw[rotate = 45] (-#1,0) -- (#1,0);
    \draw[rotate = 45] (0,-#1) -- (0, #1);
    }
}
\usepackage{amsfonts,amssymb,amsmath}

\usepackage{xcolor}

\usepackage{subcaption}

\newlength{\overwritelength}
\newlength{\minimumoverwritelength}
\newcommand{\overwrite}[3][red]{%
  \settowidth{\overwritelength}{$#2$}%
  \ifdim\overwritelength<\minimumoverwritelength%
    \setlength{\overwritelength}{\minimumoverwritelength}\fi%
  \stackrel
    {%
      \begin{minipage}{\overwritelength}%
        \color{#1}\centering\small #3\\%
        \rule{1pt}{9pt}%
      \end{minipage}}
    {\colorbox{#1!50}{\color{black}$\displaystyle#2$}}}

\usepackage{color}

\usepackage{pifont}
\usetikzlibrary{arrows.meta,positioning,calc,decorations.pathmorphing}

\tikzset{
  snakeline/.style = {->,thick, decorate, decoration={pre length=0.2cm, post length=0.2cm, snake, amplitude=.4mm, segment length=2mm}, cadmiumgreen},
  block/.style = {draw, fill=blue!20, minimum height=3em, minimum width=3em},
  pinstyle/.style={pin edge={to-,thin,black}},
}

\usetikzlibrary{shapes.misc}

\tikzset{cross/.style={cross out, draw=black, minimum size=2*(#1-\pgflinewidth), inner sep=0pt, outer sep=0pt},
cross/.default={1pt}}

\makeatletter
\newcommand \EmptyArgParse [2]
  {%
    \if\relax\detokenize{#2}\relax
      \def\ProcessedArgument{#1}%
    \else
      \def\ProcessedArgument{#2}%
    \fi
  }
\NewDocumentCommand \mathcircled { >{\EmptyArgParse{red}}O{} O{circle} m }
  {%
    \mathpalette{\mathcircled@b{#1}{#2}}{#3}%
  }
\newcommand\mathcircled@b[4]
  {%
    \tikz[baseline=(math.base)]
      \node[draw,color=#1,#2,inner sep=1pt] (math) {$\m@th#3#4$};%
  }
\makeatother

\title{\LARGE \bf Convection-Enabled Boundary Control of a 2D Channel Flow}

\author{M. C. Belhadjoudja, M. Krsti\'c, E. Witrant
\thanks{M. C. Belhadjoudja is with Universit\'e Grenoble Alpes, CNRS, Grenoble-INP, GIPSA-lab, F-38000, Grenoble, France (e-mail: mohamed.belhadjoudja@gipsa-lab.fr).}
\thanks{
M. Krsti{\'c} is with the Department of Mechanical and Aerospace Engineering, University of California San Diego, 92093
San Diego, USA, (e-mail: krstic@ucsd.edu).}
\thanks{E. Witrant is with Universit\'e Grenoble Alpes, CNRS, Grenoble-INP, GIPSA-lab, F-38000, Grenoble, France, and the Departement of Mechanical Engineering, Dalhousie University, Halifax B3H 4R2, Nova Scotia, Canada, (e-mail: emmanuel.witrant@gipsa-lab.fr).}}

\begin{document}

\maketitle

\begin{abstract}
Nonlinear convection, the source of turbulence in fluid flows, may hold the key to stabilizing turbulence by solving a specific cubic polynomial equation. We consider the incompressible Navier-Stokes equations in a two-dimensional channel. The tangential and normal velocities are assumed to be periodic in the streamwise direction. The pressure difference between the left and right ends of the channel is constant. Moreover, we consider no-slip boundary conditions, that is, zero tangential velocity, at the top and bottom walls of the channel, and normal velocity actuation at the top and bottom walls. We design the boundary control inputs to achieve global exponential stabilization, in the $L^2$ sense, of a chosen Poiseuille equilibrium profile for an arbitrarily large Reynolds number. The key idea behind our approach is to select the boundary controllers such that they have zero spatial mean (to guarantee mass conservation) but non-zero spatial cubic mean. We reveal that, because of convection, the time derivative of the $L^2$ energy of the regulation error is a cubic polynomial in the cubic mean of the boundary inputs. Regulation is then achieved by solving a specific cubic equation, using the Cardano root formula. The results are illustrated via a numerical example.
\end{abstract}

\section{Introduction} \label{intro}

The behavior of incompressible fluids (such as water) is governed by the incompressible Navier-Stokes equations. Within these equations, two critical terms influence fluid stability: nonlinear convection and diffusion. Nonlinear convection tends to enhance fluctuations and foster turbulence, while diffusion works to smooth out variations and stabilize the flow. The diffusion term in the Navier-Stokes equations is scaled by the viscosity, which is inversely proportional to the Reynolds number. Thus, the interplay between convection and diffusion, mediated by the Reynolds number, determines the overall stability of the fluid flow. In the vast majority of fluid engineering applications, controlling turbulence is of utmost importance due to its significant impact on process efficiency, energy consumption, and overall system performance and reliability. 

A classical benchmark problem in the fluid flow control community for developing turbulence control methods is the regulation of channel flows toward steady Poiseuille profiles via boundary control. A common approach is to linearize the flow equations around the desired steady state, and to develop local control strategies. Linearization is especially relevent given the chaotic nature of the Navier-Stokes equations. Indeed, small perturbations in a laminar flow, such as a Poiseuille profile, can cause transition to turbulence in the absence of an adequate control action. By linearizing the equations, one can control these perturbations and eventually prevent the onset of turbulence. Results in this direction are in \cite{NS_book}. Other significant contributions utilize reduced-order models of the linearized Navier-Stokes equations, such as in \cite{reduced_linear1,reduced_linear2,bewley1}. These approaches consist in approximating the linearized Navier-Stokes equations, which are partial differential equations (PDEs), by a finite set of ordinary differential equations (ODEs), written either in state-space form or in the frequency domain. Subsequently, standard techniques from finite-dimensional linear control theory can be applied for control design. The approximation step can be achieved, for example, through spatial discretization methods from numerical analysis. While these approaches greatly simplify the design process, their results may not fully extend to the original PDEs due to the inherent approximations in reduced-order models. In contrast, optimal control theory offers a powerful alternative for turbulence control \cite{optimal_1,optimal_2}. In optimal control, a cost functional is defined with the aim of finding a control input that minimizes (locally) this functional. However, while optimal control is effective at locally minimizing a chosen cost functional, it may not guarantee asymptotic regulation towards the desired steady state. Moreover, in optimal control, the control input is not obtained in closed-form, and one needs either to linearize the equations or to approach numerically the controller via computationally expensive techniques. We refer the reader to \cite{review1,review2,review3} for reviews and further details. 

On the other hand, global stabilization is essential, as it directly addresses the full Navier-Stokes equations rather than their linearized form. Local control strategies typically assume that the initial velocity field is \textit{close enough} to the desired Poiseuille equilibrium profile. However, this assumption may not be valid in practice, such as when control is applied after turbulence has developed. The problem of global stabilization is addressed in \cite{NS1,NS2}, but under the assumption of sufficiently small Reynolds number. To the best of our knowledge, the problem of global stabilization of channel flows governed by the full nonlinear Navier-Stokes equations, for arbitrarily large Reynolds numbers, remains open. This problem is the central focus of our work. More precisely, we consider a two-dimensional channel flow governed by the nonlinear incompressible Navier-Stokes equations, at arbitrarily high Reynolds number. We assume periodicity in the velocity field along the streamwise direction, with a constant pressure difference between the channel's left and right ends. At the top and bottom walls, the tangential velocity is zero, while normal velocity actuation is applied through blowing/suction of fluid. We design the boundary control inputs to achieve global exponential regulation, in the $L^2$ sense, and at a chosen decay rate, of any desired Poiseuille equilibrium profile. The key to our result is to exploit the nonlinear term in the Navier-Stokes equations: convection. Specifically, we reveal, through $L^2$ energy estimates on the regulation error, that convection serves as a stabilizing term when: the normal velocity at the top wall is of opposite sign to that at the bottom wall, the cubic spatial average of the normal velocity at the top and bottom walls is non-zero, and the spatial average of the normal velocity at the top and bottom walls is zero to ensure mass conservation. We show that the time derivative of the $L^2$ energy of the regulation error is bounded by a cubic polynomial in the cubic spatial average of the boundary inputs. The greater the cubic spatial average of the normal velocity at the walls, the stronger the stabilizing effect of convection. Exponential regulation at a desired decay rate is then achieved by solving a specific cubic equation using the Cardano root formula.  This result extends the \textit{Cardano-Lyapunov formula}, developed for scalar-valued convective PDEs in \cite{cardano_lyapu}, to the nonlinear two-dimensional Navier-Stokes equations. Our primary tool is the Lyapunov approach, which has already demonstrated success in certain fluid systems \cite{tank,piston1,antarctic2}.

\textbf{Notation.} The divergence of a vector-valued function $\textbf{W}(x,y,t) = (U(x,y,t),V(x,y,t))^{\top}$ with respect to $(x,y)$ is defined by $\nabla \cdot \textbf{W} := U_x + V_y$. We say that $\textbf{W}$ is divergence-free if $\nabla \cdot \textbf{W} = 0$. The Laplacian of $\textbf{W}$ is $\Delta \textbf{W} := U_{xx}+V_{yy}$. The gradient of a scalar field $P(x,y,t)$ with respect to $(x,y)$ is the vector $\nabla P := (P_x,P_y)^{\top}$. Finally, we write $(\textbf{W}\cdot \nabla)\textbf{W} := (UU_x+VU_y,UV_x+VV_y)^{\top}$. 

\section{Problem statement and main result}\label{sec3}
\subsection{Problem statement}
We consider a channel flow described by the incompressible Navier-Stokes equations 
\begin{equation}\label{NSE}
\left\lbrace
\begin{aligned}
\textbf{W}_t  - \frac{1}{R} \Delta \textbf{W}+(\textbf{W}\cdot \nabla)\textbf{W}+\nabla P &= 0, \\
\nabla \cdot \textbf{W} &= 0, 
\end{aligned}
\right.
\end{equation}
where $\textbf{W}(x,y,t) = (U(x,y,t),V(x,y,t))^\top \in \mathbb{R}^2$ is the fluid's velocity field, $P(x,y,t)\in \mathbb{R}$ is the fluid's pressure field, $R>0$ is the Reynolds number, and $(x,y,t) \in [0,L_x]\times [0,L_y]\times (0,+\infty)$, with $L_x, L_y>0$. We refer to $U$ as the tangential velocity of the fluid, and to $V$ as the normal velocity. We assume that \eqref{NSE} is subject to the no-slip boundary conditions 
\begin{align}
U(x,0,t) = U(x,L_y,t) = 0.\label{bc1_NSE}
\end{align}
Moreover, we consider periodic boundary conditions in the streamwise direction, namely,
\begin{align}
U(0,y,t) &= U(L_x,y,t), \label{perio1}\\
V(0,y,t) &= V(L_x,y,t), \label{perio2} \\
V_x(0,y,t) &= V_x(L_x,y,t). \label{perio3} 
\end{align}
The pressure is assumed to satisfy the boundary condition 
\begin{align}
P(0,y,t) = P(L_x,y,t)+aL_x, \label{pressure_b}
\end{align}
where $a\geq 0$ is constant. Finally, we consider normal velocity actuation 
\begin{align}
V(x,0,t) &= -V(x,L_y,t) =F(x)\psi(t), \label{control1}
\end{align}
where $(F,\psi)$ are functions to be designed. 

When $\psi :=0$, the system \eqref{NSE}-\eqref{control1} admits as an equilibrium the following parabolic Poiseuille profile
\begin{align}
\bar{U}(y) &:= \frac{R}{2}ay(L_y-y), \label{equ}\\
\bar{V} &:=0, \label{eqv}\\
\bar{P}(x) &:=-ax+b, \label{eqp}
\end{align}
where $b\geq0$. The objective is to design $(F,\psi)$ to achieve $L^2$ global exponential stabilization of $(\bar{U},\bar{V})$, for any value of $R$. Strictly speaking, for a desired decay rate $\alpha >0$, we aim at achieving
\begin{align}
E(t) \leq E(0)e^{-\alpha t} \,,\quad t\geq 0, \label{exp_stab}
\end{align}
where 
\begin{align}
E &:= \frac{1}{2}\int_{0}^{L_x} \int_{0}^{L_y}\big[u(x,y)^2+v(x,y)^2 \big]\ dxdy, \label{lyapu_2} \\
u&:= U-\bar{U}, \label{u} \\
v &:= V-\bar{V}. \label{v} 
\end{align}

\subsection{Main result}

We prove in this paper the following result. 
\begin{theorem}\label{thm1}
Consider the system \eqref{NSE} subject to the boundary conditions \eqref{bc1_NSE}-\eqref{control1}. Let $(\bar{U},\bar{V},\bar{P})$ be the parabolic Poiseuille equilibrium profile defined in \eqref{equ}-\eqref{eqp}, and $E$ be the $L^2$ energy of the regulation error defined in \eqref{lyapu_2}, with $(u,v)$ the deviation variables in \eqref{u}-\eqref{v}. Select $(F,\psi)$ as follows
\begin{itemize}
    \item $F$ is any function such that $F(0)=F(L_x)$, $F'(0)=F'(L_x)$, 
\begin{align}
\int_{0}^{L_x}F(x)\ dx = 0, \ \ \text{and} \ \ \int_{0}^{L_x}F(x)^3\ dx \neq 0.
\end{align}
\item $\psi$ is given by 
\begin{align}
\psi := -\frac{\Psi}{\sqrt[3]{\int_{0}^{L_x}F(x)^3\ dx}}, \label{pPsi}
\end{align}
where 
\begin{align}
&\Psi := \sqrt[3]{-\frac{q}{2} + \sqrt{\frac{q^2}{4}+\frac{\beta^3}{27}}}+\sqrt[3]{-\frac{q}{2} - \sqrt{\frac{q^2}{4}+\frac{\beta^3}{27}}}, \label{cardano_thm} \\
&\beta := 
\frac{\int_{0}^{L_x}F(x)\big[p(x,0)+p(x,L_y)\big]\ dx}{\sqrt[3]{\int_{0}^{L_x}F(x)^3\ dx}}, \label{beta_thm} \\ 
&p := 
P - \bar{P}, \label{p} \\
&q := 
\Gamma + \frac{2\sqrt{3}}{9}\big|\beta\big|^{\frac{3}{2}}, \label{q_thm} \\
&\Gamma := 
\alpha E + \bigg|\int_{0}^{L_x}\int_{0}^{L_y}u\bar{U}'v\ dxdy\bigg|. \label{g_thm}
\end{align} 
\end{itemize}
Then, along any strong solution $(U,V,P)$ to \eqref{NSE}-\eqref{control1}, we have $E(t)\leq E(0)e^{-\alpha t}$ for all $t\geq 0$.
\end{theorem}

To implement the control law in Theorem \ref{thm1}, one needs to select a periodic function $F$ with zero mean but non-zero cubic mean. An example of such a function is given below. 
\begin{example}
Let $\theta \in \mathbb{R}^*$ and $\epsilon \in (0,L_x)$ with $\epsilon \neq L_x/2$ and $\epsilon \neq L_x/4$. Let $F:[0,L_x]\to \mathbb{R}$ be defined as follows 
\begin{equation}
F(x) := 
\left\lbrace 
\begin{aligned}
&\frac{\theta (2\epsilon - L_x)}{2\epsilon} \quad &&x\in [0,\epsilon)\cup [L_x-\epsilon,L_x], \\
&\ \ \ \ \theta \quad &&x\in [\epsilon,L_x-\epsilon).
\end{aligned}
\right. \label{f_def}
\end{equation}
Then $F(0)=F(L_x)=\theta (2\epsilon - L_x)/2\epsilon$, $F'(0)=F'(L_x)=0$, $\int_{0}^{L_x}F(x)\ dx=0$, and
\begin{align}
\int_{0}^{L_x}F(x)^3\ dx = \theta^3 \bigg[(L_x-2\epsilon) + \frac{(2\epsilon - L_x)^3}{4\epsilon^2}\bigg] \neq 0.  \label{cubic_mean_value}
\end{align}
\end{example}
\begin{remark}
It is of interest to compare our control strategy, which is based on a Lyapunov-type analysis of the control system \eqref{NSE}-\eqref{control1}, with a popular control strategy in the fluid flow control community, that consists of periodically blowing and sucking fluid at the top and bottom walls of the channel \cite{periodic1,periodic2}. The latter strategy, which is open-loop, has been shown numerically to be successful for turbulence attenuation in some cases. The control takes the form of a sinusoidal function $V(x,0,t)=-V(x,L_y,t) := -2A \cos(\omega (x -ct))$, where $A$, $\omega$ and $c$ are tuned experimentally. There are two fundamental differences between our controller and the sinusoidal control input previously mentioned. First, our controller incorporates feedback, which allows us to guarantee global regulation results. Second, although our controller has a zero spatial mean, its spatial cubic mean is non-null. This non-zero cubic mean property is the key to exploit non-linear convection for stabilization. That said, specialists in the fluid flow control community who use open-loop strategies might benefit from exploring control inputs with non-zero cubic means. Such functions could potentially improve turbulence attenuation due to the resulting stabilizing effect of non-linear convection.
\end{remark}

\section{Proof of the Main Result}\label{sec4}
\subsection{Error dynamics}
The first step of the proof consists of writing the system of PDEs describing the evolution of the deviation variables $(u,v,p)$. According to \cite[Equation 19]{NS1}, it is given by
\begin{equation}
\begin{aligned}
&u_t-\frac{1}{R}\Delta u + uu_x + vu_y + \bar{U}u_x + \bar{U}'v+p_x = 0, \\
&v_t-\frac{1}{R}\Delta v + uv_x + vv_y + \bar{U}v_x +p_y = 0, \\ 
&u_x+v_y = 0. \label{Error_NS}
\end{aligned}
\end{equation}
Moreover, system \eqref{Error_NS} is subject to the following set of boundary conditions
\begin{align}
u(x,0,t) &= u(x,L_y,t) = 0, \label{err_bo} \\
u(0,y,t) &= u(L_x,y,t), \label{err_b1} \\
v(0,y,t) &= v(L_x,y,t), \label{err_b2} \\
v_x(0,y,t) &= v_x(L_x,y,t), \label{err_b3} \\
p(0,y,t) &= p(L_x,y,t), \label{err_b4} \\
v(x,0,t) &= -v(x,L_y,t) = F(x)\psi(t). \label{err_b6}
\end{align}
\subsection{$L^2$ energy estimate}
Next, we differentiate $E$ with respect to time, and we derive an upper bound on $\dot{E}$. By differentiating $E$ along the strong solutions to \eqref{Error_NS}-\eqref{err_b6}, we find 
\begin{align}
\dot{E} =&~ \int_{0}^{L_x}\int_{0}^{L_y}\big[uu_t + vv_t\big]\ dxdy \nonumber \\
=&~\int_{0}^{L_x}\int_{0}^{L_y}u\bigg[\frac{1}{R}\Delta u - uu_x - vu_y - \bar{U}u_x- \bar{U}'v \nonumber \\
&~-p_x\bigg]\ dxdy+\int_{0}^{L_x}\int_{0}^{L_y}v\bigg[\frac{1}{R}\Delta v - uv_x \nonumber \\
&~- vv_y - \bar{U}v_x - p_y\bigg]\ dxdy. \label{foo}
\end{align}
\subsubsection{\underline{Contribution of diffusion}}
Using integration by parts, we obtain 
\begin{align}
\int_{0}^{L_x}\int_{0}^{L_y} u\Delta u dxdy =&~ - \int_{0}^{L_x}\int_{0}^{L_y} \big[u_x^2+u_y^2\big] dxdy \nonumber \\
&~ +\int_{0}^{L_y}\big[uu_x]_{x=0}^{x=L_x}\ dy\nonumber \\
&~+ \int_{0}^{L_x}\big[uu_y]_{y=0}^{y=L_y}\ dx. \label{laplace1}
\end{align}
Since $u(x,0)=u(x,L_y)=0$, then $\int_{0}^{L_x} \big[uu_y\big]_{y=0}^{y=L_y}\ dx = 0$. Next, using the fact that $(u,v)$ is divergence-free, we have $u_x(0,y) = -v_y(0,y)$ and $u_x(L_x,y) = -v_y(L_x,y)$. On the other hand, by differentiating both sides of \eqref{err_b2} with respect to $y$, we find $v_y(0,y)=v_y(L_x,y)$. As a consequence, we have $u_x(0,y)=u_x(L_x,y)$. Using the periodicity of both $u$ and $u_x$ in the streamwise direction, we conclude that $\int_{0}^{L_y}\big[uu_x\big]_{x=0}^{x=L_x}\ dy = 0$. As a result,
\begin{align}
\int_{0}^{L_x}\int_{0}^{L_y} u\Delta u \ dxdy = -\int_{0}^{L_x}\int_{0}^{L_y} \big[u_x^2+u_y^2\big]\ dxdy. \label{important1}
\end{align}
Similarly, we can show that 
\begin{align}
\int_{0}^{L_x}\int_{0}^{L_y}v\Delta v\ dxdy = -\int_{0}^{L_x}\int_{0}^{L_y}\big[v_x^2+v_y^2\big]\ dxdy. \label{important2}
\end{align}
\subsubsection{\underline{Contribution of pressure}}
Using integration by parts, note that we have 
\begin{align}
\int_{0}^{L_x}\int_{0}^{L_y}up_x\ dxdy =&~ \int_{0}^{L_y}\big[up\big]_{x=0}^{x=L_x}\ dy\nonumber \\
&~- \int_{0}^{L_x}\int_{0}^{L_y}u_xp\ dxdy. 
\end{align}
Since $u$ and $p$ are periodic in the $x$ direction, then $\int_{0}^{L_y}\big[up\big]_{x=0}^{x=L_x}\ dy = 0$. As a consequence, we have 
\begin{align}
\int_{0}^{L_x}\int_{0}^{L_y}up_x\ dxdy = -\int_{0}^{L_x}\int_{0}^{L_y}u_xp \ dxdy. \label{pressure_add1}
\end{align}
On the other hand, using integration by parts, we have 
\begin{align}
\int_{0}^{L_x}\int_{0}^{L_y}vp_y\ dxdy =&~ \int_{0}^{L_x}\big[vp\big]_{y=0}^{y=L_y}\ dx \nonumber \\
&~-\int_{0}^{L_x}\int_{0}^{L_y}v_yp\ dxdy. \label{deltap}
\end{align}
Since $v(x,0,t)=-v(x,L_y,t)=F(x)\psi(t)$, then we can rewrite \eqref{deltap} as 
\begin{align}
\int_{0}^{L_x}\int_{0}^{L_y}vp_y\ dxdy =&~ -\psi\int_{0}^{L_x}F(x)p(x,L_y)\ dx \nonumber \\
&~-\psi \int_{0}^{L_x}F(x)p(x,0)\ dx \nonumber \\
&~-\int_{0}^{L_x}\int_{0}^{L_y}v_vp\ dxdy. \label{pressure_add2}
\end{align}
By adding \eqref{pressure_add1} and \eqref{pressure_add2}, and using the fact that $u_x+v_y=0$, we conclude that 
\begin{align}
\int_{0}^{L_x}&\int_{0}^{L_y}\big[up_x+vp_y\big]dxdy \nonumber \\
=&~-\psi \int_{0}^{L_x}F(x)\big[p(x,L_y)+p(x,0)\big]dx. \label{important3}
\end{align}
\subsubsection{\underline{Contribution of convection}}
Now, we analyze the effect of the terms that come from the convection $(\textbf{W}\cdot\nabla)\textbf{W}$. Using integration by parts, note that we have
\begin{align}
\int_{0}^{L_x}\int_{0}^{L_y}u^2u_x\ dxdy = \int_{0}^{L_y}\bigg[ \frac{u^3}{3}\bigg]_{x=0}^{x=L_x}\ dy. \label{eq8}
\end{align}
Since $u$ is periodic in the $x$ direction, then \eqref{eq8} becomes
\begin{align}
\int_{0}^{L_x}\int_{0}^{L_y}u^2u_x\ dxdy = 0. \label{eq9}
\end{align}
As a consequence, we have 
\begin{align}
\int_{0}^{L_x}\int_{0}^{L_y}u\big[uu_x+vu_y\big]dxdy = \int_{0}^{L_x}\int_{0}^{L_y}uvu_ydxdy. \label{eq10}
\end{align}
On the other hand, since $(u,v)$ is divergence-free, then 
\begin{align}
u\big[(u^2)_x+(uv)_y\big] =&~ 2u^2u_x + u^2 v_y + uvu_y \nonumber \\
=&~u^2\big[u_x+v_y\big] + u\big[uu_x+vu_y\big] \nonumber \\
=&~u\big[uu_x+vu_y\big]. \label{eq11} 
\end{align}
Using the identity \eqref{eq11}, we obtain  
\begin{align}
&\int_{0}^{L_x}\int_{0}^{L_y}u[uu_x+vu_y] \ dxdy = \int_{0}^{L_x}\int_{0}^{L_y}u\big[(u^2)_x \nonumber \\
&~+(uv)_y\big]dxdy= \int_{0}^{L_y}\bigg[\frac{2u^3}{3}\bigg]_{x=0}^{x=L_x} dy\nonumber \\
&~+\int_{0}^{L_x}[u^2v]_{y=0}^{y=L_y} dx - \int_{0}^{L_x}\int_{0}^{L_y} uvu_y\ dxdy.\label{eq12}
\end{align}
Using the periodicity of $u$ in the streamwise direction, and the boundary condition $u(x,0)=u(x,L_y)=0$, we can rewrite \eqref{eq12} as 
\begin{align}
\int_{0}^{L_x}\int_{0}^{L_y}u\big[uu_x+vu_y\big]dxdy = - \int_{0}^{L_x}\int_{0}^{L_y}uvu_ydxdy. \label{convect2}
\end{align}
Combining \eqref{eq10} and \eqref{convect2}, note that we have 
\begin{align}
\int_{0}^{L_x}\int_{0}^{L_y}&u\big[uu_x+vu_y\big]\ dxdy= \nonumber \\
&~-\int_{0}^{L_x}\int_{0}^{L_y}u\big[uu_x+vu_y\big]\ dxdy,
\end{align}
which implies that 
\begin{align}
\int_{0}^{L_x}\int_{0}^{L_y}u\big[uu_x+vu_y\big]\ dxdy = 0. \label{important4}
\end{align}
Next, using integration by parts, we obtain
\begin{align}
\int_{0}^{L_x}\int_{0}^{L_y} v^2v_y \ dxdy =&~ \int_{0}^{L_x} \bigg[\frac{v^3}{3}\bigg]_{y=0}^{y=L_y} \ dx \nonumber \\
=&~ -\frac{2}{3}\psi^3\int_{0}^{L_x}F(x)^3\ dx.\label{eq17}
\end{align}
As a consequence, note that we have 
\begin{align}
\int_{0}^{L_x}\int_{0}^{L_y} v\big[uv_x+&vv_y\big]dxdy = -\frac{2}{3}\psi^3\int_{0}^{L_x}F(x)^3\ dx \nonumber \\
&~+\int_{0}^{L_x}\int_{0}^{L_y}vuv_x dxdy. \label{eq19}
\end{align}
On the other hand, since $(u,v)$ is divergence-free, then 
\begin{align}
v\big[(v^2)_y + (uv)_x \big]=&~ 2v^2v_y + v^2 u_x + vuv_x \nonumber \\
=&~v^2\big[u_x+v_y\big] + v\big[vv_y+uv_x\big] \nonumber \\
=&~v\big[vv_y+uv_x\big]. \label{eq20}  
\end{align}
Using the identity \eqref{eq20}, we obtain 
\begin{align}
&\int_{0}^{L_x}\int_{0}^{L_y} v\big[uv_x+vv_y\big]dxdy = \int_{0}^{L_x}\int_{0}^{L_y} v\big[(v^2)_y \nonumber \\
&~+ (uv)_x\big]dxdy = \int_{0}^{L_x}\left[\frac{2v^3}{3}\right]_{y=0}^{y=L_y}dx \nonumber \\
&~+ \int_{0}^{L_y}\big[v^2u\big]_{x=0}^{x=L_x}dy-\int_{0}^{L_x}\int_{0}^{L_y}v_xuv\ dxdy. \label{eq21}
\end{align}
Since $u$ and $v$ are periodic in the streamwise direction, then we can rewrite \eqref{eq21} as 
\begin{align}
\int_{0}^{L_x}&\int_{0}^{L_y} v\big[uv_x+vv_y\big]dxdy \nonumber \\
=&~ -\frac{4}{3}\psi^3\int_{0}^{L_x}F(x)^3\ dx-\int_{0}^{L_x}\int_{0}^{L_y}v_xuvdxdy. \label{eq19bis}
\end{align}
Combining \eqref{eq19} and \eqref{eq19bis}, note that we have 
\begin{align}
\int_{0}^{L_x}\int_{0}^{L_y}v_xuv\ dxdy = -\frac{1}{3}\psi^3\int_{0}^{L_x}F(x)^3\ dx. 
\end{align}
As a consequence, we obtain
\begin{align}
\boxed{\int_{0}^{L_x}\int_{0}^{L_y}v\big[uv_x+vv_y\big]dxdy = -\psi^3 \int_{0}^{L_x}F(x)^3dx} \label{important5}
\end{align}
Finally, using integration by parts, and the fact that $\bar{U}$ is independent of $x$, note that we have 
\begin{align}
\int_{0}^{L_x}\int_{0}^{L_y}u\bar{U}u_x\ dxdy = \int_{0}^{L_y}\bar{U}\bigg[\frac{u^2}{2}\bigg]_{x=0}^{x=L_x}\ dy. \label{uubarux} 
\end{align}
Since $u$ is periodic in the streamwise direction, then \eqref{uubarux} becomes
\begin{align}
\int_{0}^{L_x}\int_{0}^{L_y}u\bar{U}u_x\ dxdy = 0. \label{important6}
\end{align}
Similarly, using integration by parts, the fact that $\bar{U}$ is independent of $x$, and that $v$ is periodic in the streamwise direction, we have 
\begin{align}
\int_{0}^{L_x}\int_{0}^{L_y} v\bar{U}v_x\ dxdy = 0. \label{important7} 
\end{align}
\subsubsection{\underline{Combining the contributions}}
Using \eqref{important1}, \eqref{important2}, \eqref{important3}, \eqref{important4}, \eqref{important5}, \eqref{important6}, and \eqref{important7}, we find   
\begin{align}
\dot{E} \leq&~ -\int_{0}^{L_x}\int_{0}^{L_y}u\bar{U}'v\ dxdy \nonumber \\
&~-\bigg[\int_{0}^{L_x}F(x)\big[p(x,0)+p(x,L_y)\big]\ dx\bigg] \psi \nonumber \\
&~-\bigg[\int_{0}^{L_x}F(x)^3\ dx\bigg]\psi^3. \label{edot_ineq2}
\end{align} 
\subsection{The cubic equation}
We perform the change of variable
\begin{align}
\Psi (t) := -\left[\sqrt[3]{\int_{0}^{L_x}F(x)^3\ dx}\right]\psi(t). 
\end{align}
We can rewrite \eqref{edot_ineq2} as 
\begin{align}
\dot{E}\leq -\int_{0}^{L_x}\int_{0}^{L_y}u\bar{U}'v\ dxdy + \beta(p)\Psi + \Psi^3,
\end{align}
where $\beta(q)$ is defined in \eqref{beta_thm}. Consider now the cubic equation 
\begin{align}
\Psi^3+\beta(p)\Psi + q(u,v,p) = 0,\label{cubic_eq_proof}
\end{align}
where $q(u,v,p)$ is defined in \eqref{q_thm}-\eqref{g_thm}. If $\Psi$ is a real root of \eqref{cubic_eq_proof}, then 
\begin{align}
\dot{E}\leq&~ -\int_{0}^{L_x}\int_{0}^{L_y}u\bar{U}'v\ dxdy - q(u,v,p) \nonumber \\
\leq &~ - \int_{0}^{L_x}\int_{0}^{L_y}u\bar{U}'vdxdy - \bigg|\int_{0}^{L_x}\int_{0}^{L_y}u\bar{U}'vdxdy\bigg| \nonumber \\
&~ - \frac{2\sqrt{3}}{9}\big|\beta(p)\big|^{\frac{3}{2}} - \alpha E \nonumber \\
\leq &~ - \alpha E, 
\end{align}
which would allow us to conclude on global exponential stability at the decay rate $\alpha$. The discriminant of \eqref{cubic_eq_proof} is 
\begin{align}
\frac{q^2}{4}+\frac{\beta^3}{27} = \frac{1}{4}\bigg[\Gamma^2+\frac{4\sqrt{3}}{9}\big|\beta\big|^{\frac{3}{2}}\Gamma \bigg] + \frac{1}{27}\bigg[\big|\beta|^3-\beta^3\bigg], 
\end{align}
where $\Gamma$ is defined in \eqref{g_thm}. Note that $\Gamma > 0$ if $E>0$. As a consequence, $q^2/4+\beta^3/27 >0$ as long as $E\neq 0$, i.e. as long as $(u,v)\neq 0$. Therefore, when $(u,v)\neq 0$, the cubic equation \eqref{cubic_eq_proof} admits a unique real root which is given by the Cardano-Lyapunov formula \eqref{cardano_thm}. Now, if $(u,v)=0$, then $q=\beta = 0$. Indeed, when $(u,v)=0$, we have, according to \eqref{Error_NS}, $p_x=0$ and $p_y=0$. It implies that $p$ is constant in $(x,y)$. As a result, note that we have 
\begin{align}
\int_{0}^{L_x}F(x)&\big[p(x,0)+p(x,L_y)\big]\ dx = \nonumber \\ 
&~\big[p(x,0)+p(x,L_y)\big]\int_{0}^{L_x}F(x)\ dx. \label{beta_zero}
\end{align}
Since $F$ has zero mean, then
\begin{align}
\int_{0}^{L_x}F(x)&\big[p(x,0)+p(x,L_y)\big]\ dx = 0,
\end{align}
which implies that $\beta = 0$. On the other hand, $E=0$, and $\int_{0}^{L_x}\int_{0}^{L_y}u\bar{U}'v\ dxdy = 0$, which implies that $q=0$. Therefore, when $(u,v)=0$, the cubic equation \eqref{cubic_eq_proof} admits the unique root $\Psi=0$, which is given by \eqref{cardano_thm}.

\begin{figure}
\centering
\begin{minipage}[b]{.5\textwidth}
\includegraphics[width=\textwidth]{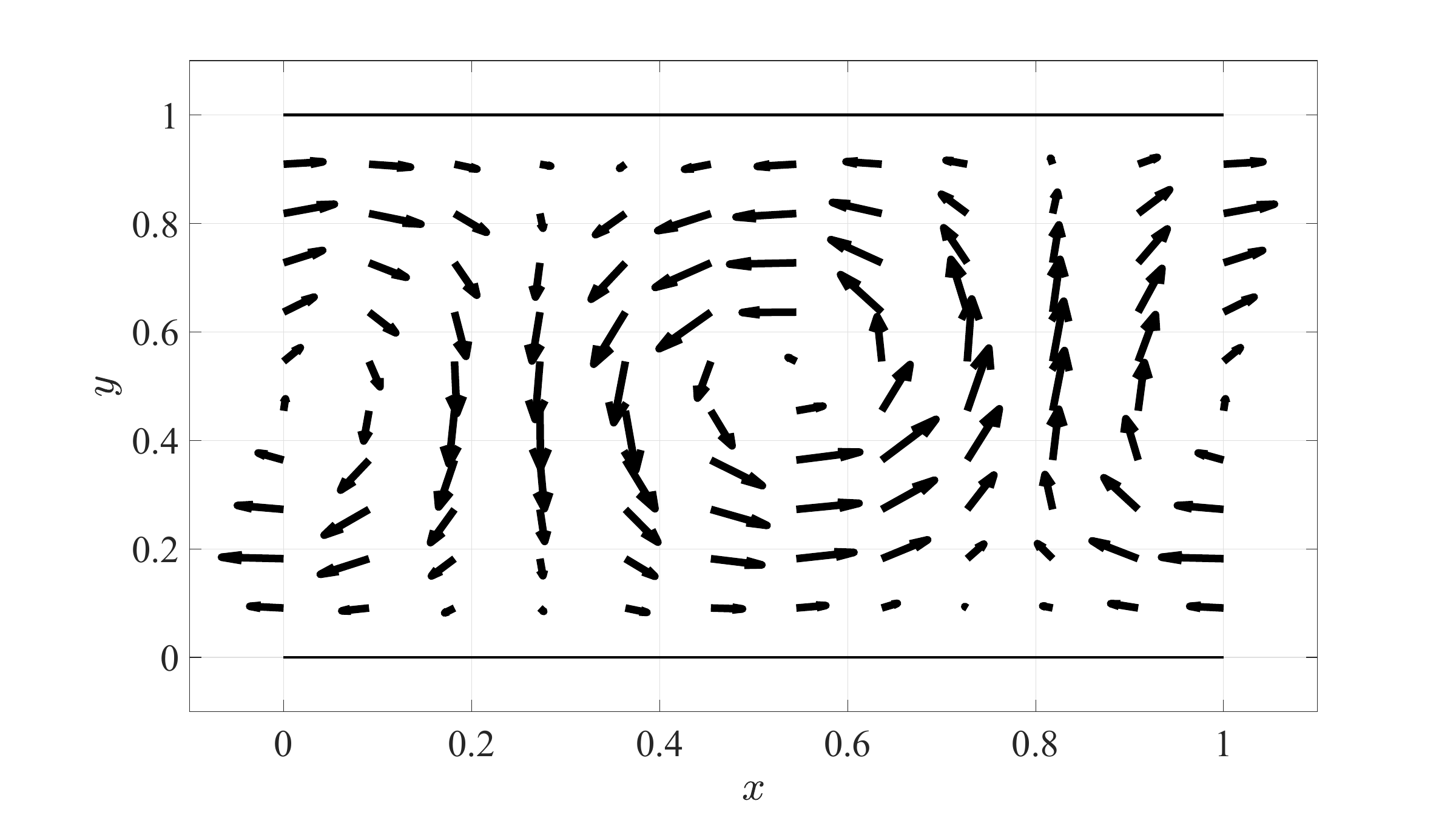}
\end{minipage}
\begin{minipage}[b]{.5\textwidth}
\includegraphics[width=\textwidth]{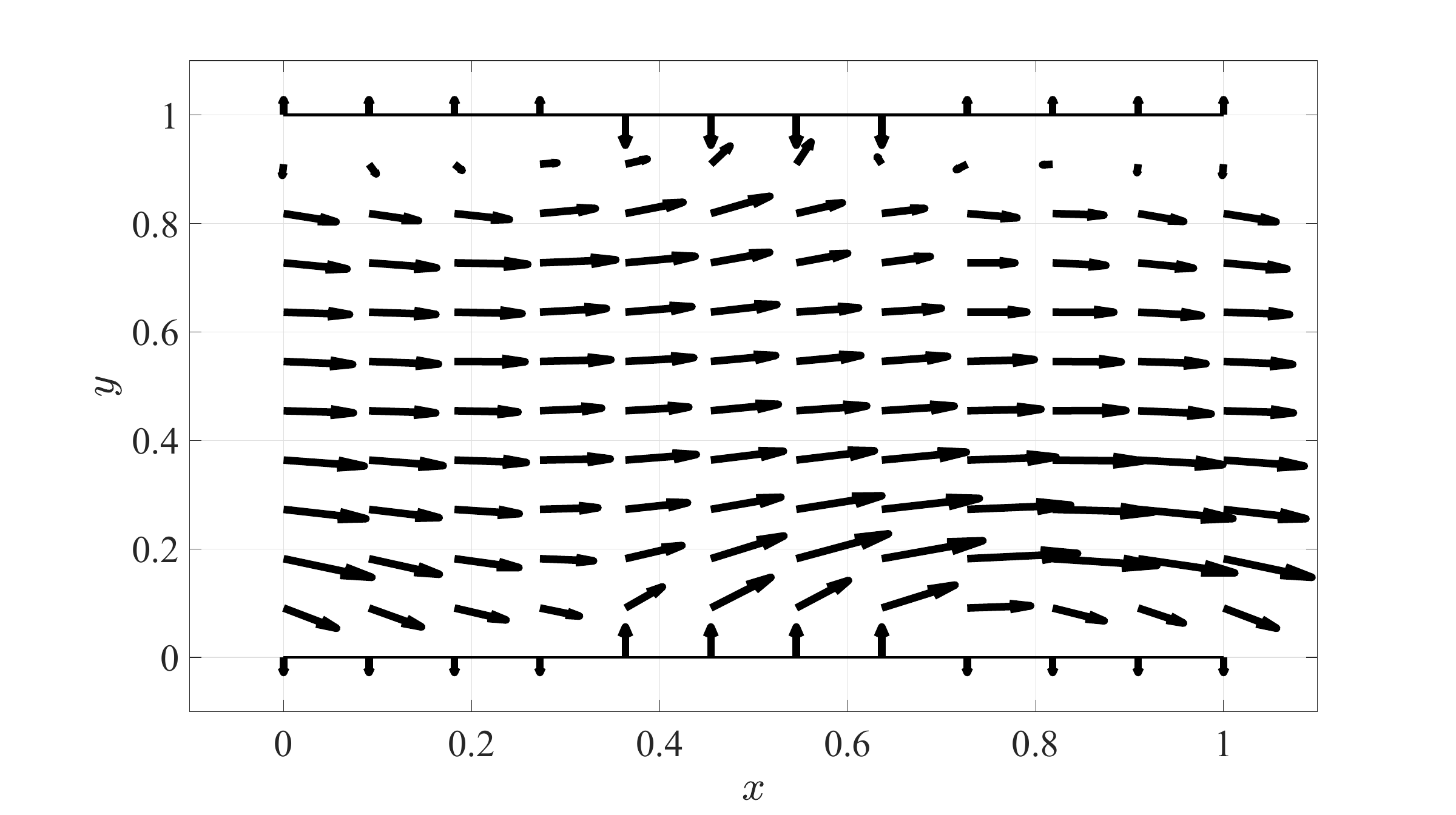}
\end{minipage}
\caption{Open-loop response (top) and closed-loop response (bottom) at Reynolds number $R:=5\times 10^4$. The energy of the regulation error in closed-loop is $10\%$ of the energy of the regulation error in open-loop at the final iteration.}\label{fig1}
\end{figure}

\section{Numerical Example} \label{simu}
We illustrate Theorem \ref{thm1} by a numerical example. Inspired by \cite{RBF_NS}, we employ a (RBF)-based decomposition of the flow variables in space. For time advancement, we use an Euler implicit scheme. To handle the non-linear convective terms, we delay the flow-variables by one time step. That is, at time iteration $n+1$, we replace $U^{n+1}U_x^{n+1}$ by $U^nU_x^{n+1}$ (and similarly for the other non-linear terms); see \cite{deblois}. The control input is delayed by one time step. The integral terms are approximated using the MATLAB \textit{trapz} function. We use multiquadrics RBFs, with the shape parameter $c=0.11$. We construct a regular mesh of size $12\times 12$ on a 2D channel of size $[0,1]\times [0,1]$. The time step is $\delta t := 0.2$. The same numerical parameters are used to simulate both the closed-loop and the open-loop responses. The Reynolds number is $R:=5\times 10^4$, which is eight times greater than the critical value $R=5772$, corresponding to the loss of linear stability. The equilibrium profile is given by \eqref{equ}-\eqref{eqp} with $a=4/R$ and $b=0$. The initial condition is $U(x,y,0):=0$, $V(x,y,0):=\sin (4\pi x)$, and $P(x,y,0):=-\frac{4}{R}x$. The control gain is $\alpha := 1$, and the function $F$ is chosen according to \eqref{f_def}, with $\theta :=1$ and $\epsilon := 1/3$. We run the simulations for $250$ iterations. The velocity profiles $(U,V)$ for the open-loop and closed-loop responses, at the final iteration, are shown in Figure \ref{fig1}.

\section{Conclusion and Prelude} \label{sec5}
In this paper, we introduced the idea that convection, the source of turbulence, may be the key to channel flow control. The question remains: \textit{Can we develop a theory based on this idea?} This work may mark the beginning of a long and exciting journey in exploring the potential of convection-enabled flow control. A first step on this path would be to extend our results to three-dimensional channels and conduct extensive numerical simulations using advanced solvers. 
 
\bibliography{biblio}
\bibliographystyle{ieeetr}

\end{document}